\documentclass{emulateapj}

\usepackage{subfigure}
\usepackage{url}
\usepackage{hyperref}
\usepackage{datetime}
\usepackage{longtable}
\usepackage{natbib}
\usepackage{amsmath}
\usepackage{ulem}
\usepackage{bm}

\newcommand{\noprint}[1]{}

\usepackage{color}

\newcommand{\itk}{{\it Kepler}}

\newcommand{\rstar}{{$R_\star$}}
\newcommand{\rhostar}{{$\rho_\star$}}

\newcommand{\rsun}{{R$_\odot$}}
\newcommand{\msun}{{M$_\odot$}}
\newcommand{\mjup}{{M$_\textrm{Jup}$}}
\newcommand{\rjup}{{R$_\textrm{Jup}$}}
\newcommand{\LA}{{LHS\,6343\,A}}
\newcommand{\LB}{{LHS\,6343\,B}}
\newcommand{\LC}{{LHS\,6343\,C}}
\newcommand{\LHS}{{LHS\,6343}}

\newcommand{\mstar}{{$M_\star$}}
\newcommand{\logg}{{log(g)}}

\begin{document}
\title{Characterizing the Cool KOI\MakeLowercase{s}. VII. \\ Refined Physical Properties of the Transiting Brown Dwarf \LC}

\author{Benjamin T. Montet\altaffilmark{1,2}, John Asher Johnson\altaffilmark{2}, Philip S. Muirhead\altaffilmark{3},
Ashley Villar\altaffilmark{2}, Corinne Vassallo\altaffilmark{4}, Christoph Baranec\altaffilmark{5}, Nicholas M. Law\altaffilmark{6},
Reed Riddle\altaffilmark{1}, Geoffrey W. Marcy\altaffilmark{7}, Andrew W. Howard\altaffilmark{8}, Howard Isaacson\altaffilmark{7}}

\email{btm@astro.caltech.edu}

\altaffiltext{1}{Cahill Center for Astronomy and Astrophysics, California Institute of Technology, 1200 E. California Blvd., MC 249-17, Pasadena, CA 91106, USA}
\altaffiltext{2}{Harvard-Smithsonian Center for Astrophysics, 60 Garden
Street, Cambridge, MA 02138, USA}
\altaffiltext{3}{Department of Astronomy, Boston University, 725 Commonwealth Avenue, Boston, MA, 02215, USA}
\altaffiltext{4}{Department of Aerospace Engineering and Engineering Mechanics, The University of Texas at Austin, 210 East 24th Street, Austin, TX 78712, USA}
\altaffiltext{5}{Institute for Astronomy, University of Hawai`i at M\={a}noa, Hilo, HI 96720, USA}
\altaffiltext{6}{Department of Physics and Astronomy, University of North Carolina at Chapel Hill, Chapel Hill, NC 27599, USA}
\altaffiltext{7}{Department of Astronomy, University of California, Berkeley, CA 94720, USA}
\altaffiltext{8}{Institute for Astronomy, University of Hawaii, 2680 Woodlawn Drive, Honolulu, HI 96822, USA}

\date{\today, \currenttime}

\begin{abstract}
We present an updated analysis of \LHS, a triple system in the \itk{} field which consists of a brown dwarf transiting one member of a widely-separated M+M binary system. 
By analyzing the full \itk{} dataset and 34 Keck/HIRES radial velocity observations, we measure both the observed transit depth and Doppler semiamplitude to 0.5\% precision. With Robo-AO and Palomar/PHARO adaptive optics imaging as well as TripleSpec spectroscopy, we measure a model-dependent mass for \LC{} of $62.1 \pm 1.2$ \mjup{} and a radius of $0.783 \pm 0.011$ \rjup. 
We detect the secondary eclipse in the \itk{} data at $3.5\sigma$, measuring $e \cos \omega = 0.0228 \pm 0.0008$.
We also derive a method to measure the mass and radius of a star and transiting companion directly, without any direct reliance on stellar models.
The mass and radius of both objects depend only on the orbital period, stellar density, reduced semimajor axis, Doppler semiamplitude, eccentricity, and inclination, as well as the knowledge that the primary star falls on the main sequence. With this method, we calculate a mass and radius for \LC{} to a precision of 3\% and 2\%, respectively.
\end{abstract}


\maketitle

\section{Introduction}
\label{sec:intro}

The growth of brown dwarf astronomy has closely mirrored that of exoplanetary astronomy.
Although \citet{Latham89} discovered a likely brown dwarf candidate, the first confirmed detection of a brown dwarf was announced two months before the announcement of the first exoplanet orbiting a main sequence star \citep{Rebolo95, Mayor95}. 
That same year also saw the discovery of the first brown dwarf orbiting a stellar-mass companion \citep{Nakajima95}. 
Today, more than 2,000 brown dwarfs have been discovered.
The majority of these substellar objects have no detected companions, so characterization is often limited to spectroscopic observations.
In these cases, the atmosphere of the brown dwarf can be extensively studied \citep[e.g.][]{Burgasser14, Faherty14}, but its physical parameters, including mass and radius, cannot be measured directly.

When a brown dwarf with a gravitationally bound companion is detected, detailed characterization of its physical properties is possible. 
Radial velocity (RV) surveys have produced a significant number of brown dwarf candidates with minimum mass determinations \citep[e.g.][]{Patel07}. 
Astrometric monitoring of directly imaged brown dwarf companions to stars has led to dynamical mass measurements of brown dwarfs \citep{Liu02, Dupuy09, Crepp12a}.
While there are many brown dwarfs with measured masses, radii can only be directly measured in transiting or eclipsing systems.
The first eclipsing brown dwarf system, discovered by \citet{Stassun06} in the Orion Nebula, produced the first measurement of a brown dwarf's radius and the first test of theoretical mass-radius relations.
Today, there are eleven brown dwarfs with measured masses and radii \citep{Diaz14}. 
Of this sample, eight transit a stellar-mass companion and only four are not inflated due to youth or irradiation.
If the brown dwarf is assumed to be coeval with its host star, the brown dwarf's age and metallicity can be estimated. Both properties are expected to affect the brown dwarf mass-radius relation, making observations of transiting brown dwarfs especially valuable \citep{Burrows11}. 

Recently, four brown dwarfs have been detected by the \itk{} mission \citep{Johnson11, Bouchy11b, Diaz13, Moutou13}.
Launched in 2009, the \itk{} telescope collected wide-field photometric observations of approximately 200,000 stars in Cygnus and Lyra every 30 minutes for 4 years \citep{Borucki10}. 
The mission was designed as a search for transiting planets.
As brown dwarfs have radii similar to Jupiter, brown dwarfs were also easily detected; only a few RV observations are necessary to distinguish between a giant planet and brown dwarf companion \citep[e.g.][]{Moutou13}.

The first unambiguous brown dwarf detected from \itk{} data was found in the \LHS{} system and announced by \citet[hereafter J11]{Johnson11}. 
The authors analyzed five transits of the primary star observed in the first six weeks of \itk{} data, combined with one transit observed in the Z-band with the Nickel telescope at Lick observatory and 14 RV observations with Keck/HIRES.  
The authors also obtained PHARO adaptive optics imaging data from the Palomar 200 inch telescope, imaging a companion 0.5 magnitudes fainter than the primary at a separation of 0\farcs7. 
From these observations, the authors were able to measure a mass for the brown dwarf of $62.7 \pm 2.4$ \mjup{}, a radius of $0.833 \pm 0.021$ \rjup, and a period of $12.71$ days, corresponding to a semimajor axis of $0.0804\pm0.0006$ AU.
The authors define \LA{} as the primary star, \LB{} as the widely-separated binary M dwarf, and \LC{} as the brown dwarf orbiting the A component, and note the architecture of this system is very similar to the NLTT\,41135 system discovered by \citet{Irwin10}.

Additional papers have expanded our knowledge of \LHS. 
\citet{Southworth11} re-fit the \itk{} light curve, using data through Quarter 2 from the mission. By fitting the observations using five different sets of stellar models, he attempted to reduce biases caused by any one individual stellar model. He found different models provide a consistent brown dwarf radius at the $0.08$ \rjup{} level, but found a higher mass than J11: his best fitting mass for \LC{} was $70 \pm 6$ \mjup.
\citet{Oshagh12} analyzed the lack of transit timing variations in the system, finding that any additional companions to \LA{} with an orbital period smaller than 100 days must have a mass smaller than that of Jupiter. 
With 6 quarters of \itk{} data, \citet{Herrero13} measured a photometric rotation period of $13.13 \pm 0.02$ days for \LA. 
The authors also claimed to observe spot-crossing events during the transits of \LA{}, as well as out-of-transit photometric modulation with a period consistent with the orbital period of \LC. 
\citet{Herrero14} updated this work, concluding that the out-of-transit variations are dominated by relativistic Doppler beaming.

In many of the papers about the \LHS{} system after the discovery paper, the authors assumed the physical parameters of J11. 
This is not necessarily an ideal assumption to make.
J11 used a limited dataset during their analysis.
Their photometry consisted of only six transits and 14 RVs, and they estimated the third light contribution of \LB{} by extrapolating from near-IR observations to the \itk{} bandpass.
Moreover, the derived stellar parameters in that paper were based only on photometric observations and depend strongly on the accuracy of the Padova model grids \citep{Girardi02} upon which they are based.

The conclusion of the primary \itk{} mission affords us an opportunity to reanalyze the \LHS{} system using the complete \itk{} dataset. 
Such a reanalysis enables us to better measure the brown dwarf's mass and radius. 
There are only three non-inflated brown dwarfs with both a mass and radius measured to $5\%$ or better: \LC, KOI-205 b \citep{Diaz13}, and KOI-415 b \citep{Moutou13}.  
To test theoretical brown dwarf evolutionary models, we would like to measure the masses, radii, and metallicities of these objects as precisely as possible. 
In this work, we analyze the full \itk{} dataset for this object to measure the transit profile.
We combine this light curve with additional RV observations, near-infrared spectroscopy of \LA B, and Robo-AO visible-light adaptive optics.
Without any reliance on stellar models beyond an empirical main sequence mass-radius relation, we are able to measure the mass of \LC{} to a precision of 3\% and the radius to a precision of 2\%. 
The mass and radius measurements depend only on the following parameters, all measured directly from the data: the orbital period, stellar density \rhostar, reduced semimajor axis $a/$\rstar, Doppler semiamplitude $K$, eccentricity, and inclination.
Our technique allows one to calculate the mass and radius for both members of a transiting system.
We also combine our data with the predictions for the mass of \LA{} from the Dartmouth stellar evolutionary models of \citet{Dotter08}.
These combined data enable us to measure a model-dependent mass and radius of \LC{} to better than $2\%$ each; we also measure a metallicity of the system of $0.02 \pm 0.19$ dex. 

In \textsection{2} we describe the observations used in this paper. 
In \textsection{3} we outline our data analysis pipeline. 
In \textsection{4} we present our results.
In \textsection{5} we summarize our present efforts and outline our future plans to measure the brown dwarf's luminosity.
In the Appendix, we derive the relation between transit and RV parameters and the mass and radius of both the primary and secondary companion.

This study presents, to date, the most precise mass and radius measurements of a non-inflated brown dwarf. 
Observations such as these are essential for future detailed characterization of field brown dwarfs.

\section{Observations}
\subsection{\itk\ Photometry}

The \LHS{} system (KIC 10002261, KOI-959) was part of the initial \itk{} target selection and was observed during all observing quarters in long cadence mode.
Between 22 February 2011 and 14 March 2011, the system was also observed using \itk's short cadence mode, with observations collected every 58.84876 seconds in the reference frame of the spacecraft. 
We downloaded the entire dataset from the NASA Multimission Archive at STScI (MAST). 

For both long and short cadence observations, \itk{} data consist of a postage stamp containing tens of pixels, a small number of which are combined to form an effective aperture.
The flux from all pixels in the aperture are combined to create a light curve. 
The \itk{} team defines an aperture for all targets and performs aperture photometry as a part of their Photometric Analysis (PA) pipeline, which produces a light curve from the pixel-level data \citep{Jenkins10}. 
This pipeline also removes the photometric background and cosmic rays. 

In analyzing the pipeline-generated light curve, we detected occasional anomalies during transit events, with the recorded flux systematically larger than expected.
These anomalies were also detected by \citet{Herrero13}, who attribute them to occultations of spots on \LA{} by \LC. 
The anomalies occur only in the long cadence data, and only when the transit is symmetric around one data point in the \itk{} time series, so that the central in-transit flux measurement would be expected to be significantly lower than the surrounding data points.
By investigating the pixel-level data, we find that each anomaly has been registered as a cosmic ray by the PA pipeline, and ``corrected'' to an artificially large value.

Using the pixel-level data, recorded before the cosmic ray correction in the pipeline, we removed these artificial corrections. 
We find the anomalies can be completely explained as false cosmic ray detections: there is no evidence for transit-to-transit variability in the \itk{} data.

We expect stellar granulation to induce correlated photometric variability only at a level significantly below the precision of our observations.
Correlated noise attributed to stellar granulation has been previously observed when modeling transits of companions to higher mass stars \citep[e.g.][]{Huber13} and used to derive fundamental parameters of the stars themselves \citep{Bastien13}.
Both the timescale and magnitude of the correlated noise are inversely proportional to the stellar density \citep{Gilliland10}.
For an M dwarf with a mass around 0.3\msun, we expect granulation to induce correlated noise with a period of approximately 10 seconds and an amplitude of 50 ppm \citep{Winget91}. 
Therefore, given the precision and cadence of the \itk{} observations we do not expect to observe correlated noise due to granulation in the \LHS{} system. 

We tested for correlated noise on transit timescales by calculating the autocorrelation matrix for out-of-transit sections of the data. For both long cadence and short cadence data, all off-diagonal elements have absolute values less than 0.03; we found no periodic structure to the autocorrelation matrix. 
Therefore, on transit timescales the noise can be treated as white.

We converted all times recorded by \itk{} to Barycentric Dynamical Time (TDB), not UTC, which was mistakenly recorded during the first three years of the mission. 
As a result, our times differ from those reported in the analysis of J11 by 66.184 seconds.

We then detrended the light curve to remove the effects of stellar and instrumental variability. 
For all transit events with at least four data points recorded continuously before and after the transit, we selected a region bounded by a maximum of three transit durations on either side of the nominal transit center.
If there is any spacecraft motion, such as a thruster fire or data downlink, we clipped the fitting region to not include these data.
We then fit a second-order polynomial to the out of transit flux. 
We normalized the light curve by dividing the observed flux values by the calculated polynomial. 
We repeated this procedure near the midpoint between successive transits in order to search for evidence of a secondary eclipse. 
We estimated the noise level in the data by measuring the variance observed in the out of transit segments of the data.

\subsection{Keck/HIRES Radial Velocities}

We obtained spectroscopic observations of \LHS{} using the HIgh Resolution Echelle Spectrometer (HIRES, $R \approx 48$,000) at the W. M. Keck Observatory. 
All observations were taken using the C2 decker. 
With a projected length of 14.0 arcsec, the decker enables accurate sky subtraction. 
The first four observations were obtained using a 45 minute exposure time and the standard iodine-cell setup described by \citet{Howard10}. 
Once \LC{} was identified as an transiting brown dwarf, the remaining observations were obtained with 3 minute exposure times and without the iodine cell.
For all observations, the slit was aligned along the binary axis so that light from both stars fell upon the detector. 

To measure the RV of \LA, we used \LB{} as a wavelength reference. 
We began with an iodine-free spectrum of HIP 428, oversampled onto a grid with resolution 15 m s$^{-1}$.
For each observation, we restricted our analysis to the 16 orders covered by the ``green'' CCD chip, which covers the region typically used in iodine cell analyses, as well as the first two orders covered by the ``red'' chip where telluric contamination is negligible.
From these 18 orders, we first estimated and divided out the continuum flux level following the method of \citet{Pineda13}. 
We then removed the regions of the spectrum contaminated by telluric lines.
We added to this template a shifted, scaled version of itself to represent \LB.
We varied the positions of both stars and compared to the observed spectrum of \LHS{} in order to find the maximum likelihood velocity separation between the two stars.
By assuming the relative RV of \LB{} does not change over our observing baseline, our method enables us to measure the RV of \LA{} relative to that of a stationary wavelength calibration source observed simultaneously. 

There is no evidence of orbital motion of \LB{} at the level of our RV precision.
From an observed projected separation and mass estimate we can estimate the maximum expected RV acceleration induced by a companion.
Following \citet{Torres99} and \citet{Knutson14}, the maximum RV acceleration is defined such that 
\begin{equation}
\left|{\dot{v}}\right| < 68.8 {\rm m\phantom{0}s^{-1}\phantom{0}yr^{-1}} \bigg( \frac{M_{\rm comp}}{M_{\rm Jup}}\bigg) \left(\frac{d}{\rm pc}\frac{\rho}{\rm arcsec}\right)^{-2},
\end{equation}
for a system at a distance $d$, with a companion with mass $M_\textrm{comp}$ at an angular separation $\rho$.
For a companion with a mass approximately 30\% of the Sun's and a projected separation ($d\rho$) of approximately 20 AU, we expect a maximum RV acceleration of 40 m s$^{-1}$ yr$^{-1}$.
We would only observe this RV acceleration if we happened to observe the two stars at the time of their maximum orbital separation and if their orbit was edge-on to our line of sight. 
Our RV signal is considerably larger than any effects induced by \LB; any RV acceleration over our three-year baseline is similar in size to our measurement uncertainties.

The median RV precision of our observations is 85 m s$^{-1}$.
Our RV precision is much lower ($\approx 400$ m s$^{-1}$) for the first four observations when the spectra are contaminated by the iodine cell. 
Our RV precision is also impeded when the difference between the RV of \LA{} and \LB{} is smaller than one-half of a pixel, about 500 m s$^{-1}$.

A table of our RVs is included as Table \ref{RVTable}.

\begin{deluxetable}{lcc}
\tabletypesize{\footnotesize}
\tablecolumns{3}
\tablecaption{Radial Velocities for LHS\,6343}
\tablehead{JD $- 2440000$ & RV (km s$^{-1}$) & Uncertainty (km s$^{-1}$)}
\startdata
15373.095 & 12.993 & 0.498 \\ 
15373.998 & 13.878 & 0.429 \\ 
15377.078 & 3.041 & 0.425 \\ 
15377.098 & 2.825 & 0.423 \\ 
15378.030 & -2.470 & 0.562 \\ 
15379.052 & -4.599 & 0.076 \\ 
15380.127 & -5.967 & 0.082 \\ 
15380.827 & -5.412 & 0.089 \\ 
15380.831 & -5.015 & 0.166 \\ 
15395.984 & 3.726 & 0.084 \\ 
15396.970 & 8.522 & 0.068 \\ 
15404.974 & -5.447 & 0.092 \\ 
15405.821 & -5.618 & 0.074 \\ 
15406.865 & -3.860 & 0.086 \\ 
15407.853 & -0.495 & 0.666 \\ 
15413.032 & 11.540 & 0.072 \\ 
15414.009 & 7.951 & 0.089 \\ 
15668.120 & 8.714 & 0.161 \\ 
15669.083 & 4.243 & 0.174 \\ 
15673.982 & -3.661 & 0.083 \\ 
15705.917 & 10.005 & 0.093 \\ 
15843.859 & 13.444 & 0.084 \\ 
16116.017 & -3.562 & 0.077 \\ 
16164.014 & 8.408 & 0.064 \\ 
16172.915 & 10.070 & 0.078 \\ 
16192.886 & -4.885 & 0.073 \\ 
16498.042 & -5.035 & 0.079 \\ 
16506.891 & 9.963 & 0.073 \\ 
16513.001 & -3.995 & 0.081 \\ 
16513.988 & 0.033 & 0.733 \\ 
16522.939 & -3.889 & 0.078 \\ 
16524.890 & -5.555 & 0.113 \\ 
16524.892 & -5.473 & 0.081 \\ 
16530.943 & 13.348 & 0.092  
\enddata
\label{RVTable}
\end{deluxetable}

\subsection{Visible-light Adaptive Optics Imaging}

\begin{figure*}[htbp!]
\centerline{\includegraphics[width=0.85\textwidth]{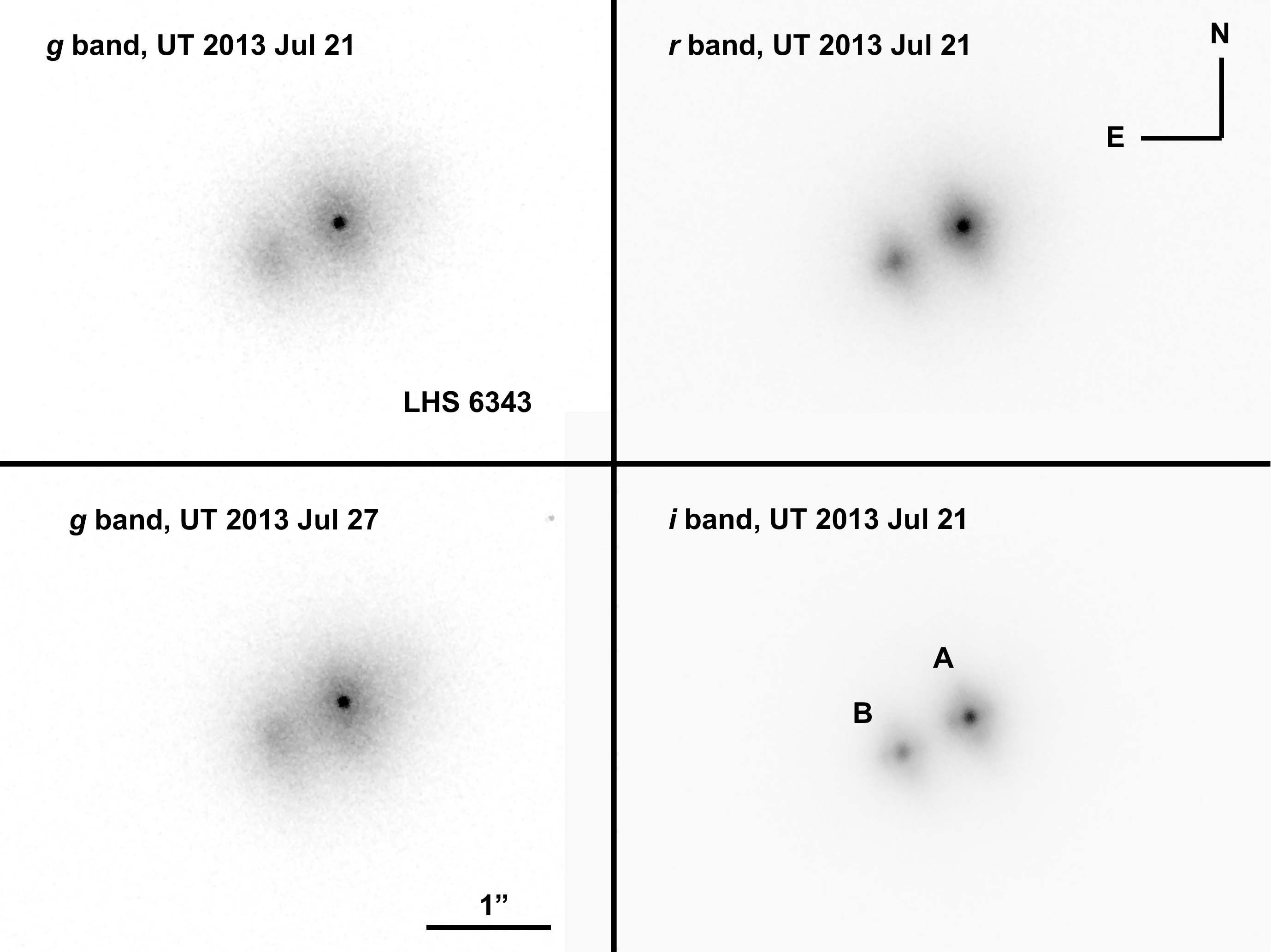}}
\caption{Robo-AO adaptive optics imaging of the \LHS{} system taken with three different bandpasses.
Both the scale and orientation are held constant across all images.
We obtained two images of the system in the $g$-band, six days apart.
We obtained a single image of the system in both the $r$- and $i$-bands.
  }
\label{AOPlot}
\end{figure*}

J11 estimated the third-light contribution of \LB{} in the \itk{} bandpass by extrapolating from JHK adaptive optics observations using the Padova model atmospheres of \citet{Girardi02}. 
To minimize any potential biases that may be induced by their reliance on stellar models, we obtained adaptive optics imaging of \LHS{} with the Robo-AO laster adaptive optics and imaging system on the Palomar Observatory 60-inch telescope \citep{Baranec14}. 
Robo-AO successfully observed thousands of KOIs; we used their standard setup \citep{Law14}. 
With SDSS $g$, $r$, and $i$ filters \citep{York00}, we imaged the system on UT 2013 21 July; we observed the system again in $g$ band on UT 2013 27 July. 
Each observation consisted of full-frame-detector readouts at 8.6 Hz for 90 seconds. 
We use 100\% of the frames during each integration.
The images were then combined using a shift-and-add processing scheme, using \LA{} as the tip-tilt star. 
At all wavelengths, we detected both \LA{} and \LB, as shown in Figure \ref{AOPlot}.
While we would be sensitive to a change in the position angle between the two M dwarfs of two degrees, we do not detect any orbital motion of \LB{} relative to \LA{} between the original Palomar/PHARO data in 2010 and these observations in 2013.

To calculate the relative flux ratio of the two stars in each bandpass, we sky-subtract our observations and measure the flux inside a 0\farcs5 aperture centered on each star. 
The point spread functions of each star are larger than the apertures, so each aperture contains light from both stars. We subtract out the contamination from each star by measuring the flux in a similar aperture on the opposite side of each star.

In our $g$-band data we observed tripling, induced when the shift-and-add processing algorithm temporarily locks on \LB{} instead of \LA.
Tripling causes the appearance of an artificial third object coaxial with the two real objects.
The third object is observed to have the same projected separation between the primary as the true secondary, at a position angle offset of 180 degrees, as discussed by \citet{Law06b}.
By measuring the flux ratios between the primary star and the two imaged companions, and defining $I_{jk} \equiv F_j/F_k$, then the true binary flux ratio $F_R$ is
\begin{equation}
F_R = \frac{2 I_{13}}{I_{12}I_{13} + \sqrt{I_{12}^2 I_{13}^2 - 4 I_{12} I_{13}}},
\end{equation}
where $F_1$ is the observed flux from the primary component, $F_2$ the observed flux from the secondary component, and $F_3$ the observed light from the tertiary, ``tripled'' component.
When $F_3 = 0$ this equation is undefined, but the asymptotic behavior is correct.

We find the third light contributions in each bandpass are given such that $\Delta g = 0.93 \pm 0.07$, $\Delta r = 0.74 \pm 0.06$, and $\Delta i = 0.57 \pm 0.05$.  
From these, we interpolate using the Dartmouth stellar models to calculate a value for the third light in the \itk{} bandpass, which encompasses roughly the $g$, $r$, and $i$ filters.
We find $\Delta K_p = 0.71 \pm 0.07$ magnitudes.
This is consistent with the extrapolation of J11, who predict a third-light in the \itk{} bandpass of $\Delta K_p = 0.74 \pm 0.10$.

\subsection{NIR Spectroscopy}

The transit light curve itself can be used to measure some properties of \LA, such as the stellar density.
Other parameters such as the stellar temperature, as well as all physical properties of \LB, can only be estimated by relying on stellar models. 
To inform the models, on UT 2012 July 05 we obtained simultaneous JHK spectroscopy with the TripleSpec Spectrograph on the 200" Hale Telescope at Palomar Observatory. 
TripleSpec is a near-infrared slit spectrograph with a resolving power ($\lambda / \Delta \lambda$) of 2700 \citep{Wilson04, Herter08}. 

Observations were collected on four positions along the slit, ABCD, to minimize the effects of hot and dead pixels on the spectrograph detector. 
Each exposure was 30 seconds long in order to achieve a signal-to-noise ratio of 60.
We then observed a nearby, rapidly rotating A0V star to calibrate absorption lines caused by the Earth's atmosphere.

\begin{figure}[htbp!]
\centerline{\includegraphics[width=0.5\textwidth]{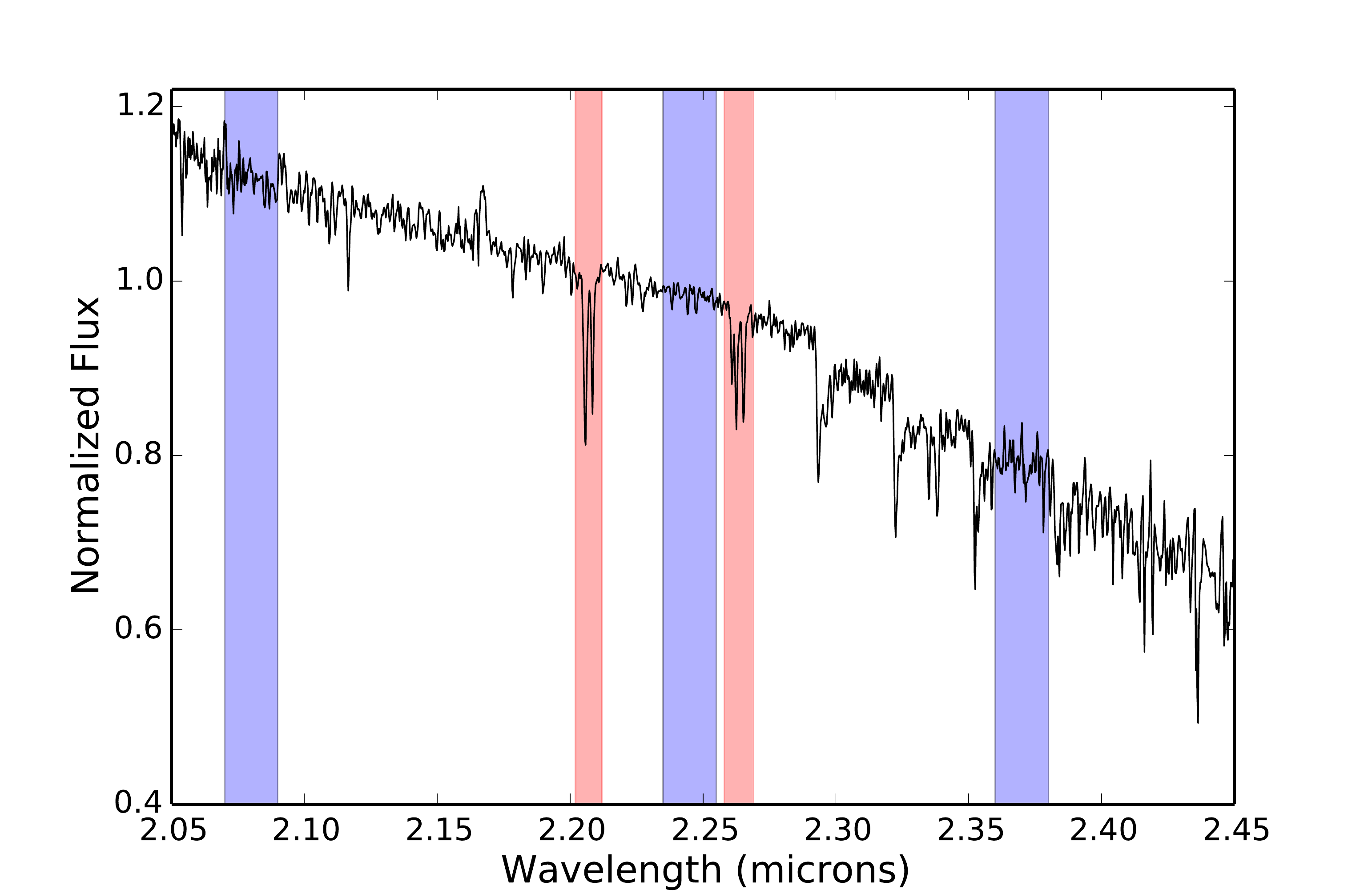}}
\caption{Combined-light K-band spectrum for the \LHS{} system.
The broad, blue shaded regions are used to derive the ``H$_2$O--K2 water index,'' as described in \textsection\ref{TripleSpec}.
The narrow, red shaded regions encompass the sodium doublet and calcium triplet.
Together, these regions have been used to develop empirical relations for the temperature and metallicity of M dwarfs \citep{RojasAyala12}.
  }
\label{TripleSpecPlot}
\end{figure}

To reduce the data, we followed the methodology of \citet{Muirhead14}, using the SpexTool reduction package of \citet{Cushing04}. 
We differenced the A and B observations and the C and D observations separately, then extracted the combined-light spectrum and combined the separate observations with SpexTool. 
To remove the system's absolute radial velocity of -46 km s$^{-1}$, we cross-correlated the spectrum with data from the IRTF spectral library \citep{Cushing05, Rayner09}, then applied an offset to the wavelength solution corresponding to the peak of the cross-correlation function. 
The result is a single spectrum displaying the combined light from \LA{} and \LB, as shown in Figure \ref{TripleSpecPlot}

\section{Data Analysis}

\subsection{Temperature and Metallicity of LHS6343\,A and B}
\label{TripleSpec}
We measured the temperature of each star following the method of \citet{RojasAyala12}, who built on the efforts of \citet{Covey10} to determine a relation between K-band spectroscopic features and the temperature and metallicity of M dwarfs. 
Specifically, Rojas-Ayala et al. define a temperature-sensitive ``H$_2$O--K2 water index," representing the water opacity between 2.07 $\mu$m and 2.38$\mu$m:
\begin{equation}
\textrm{H}_2\textrm{O--K}2 = \frac{\langle\mathcal{F}(2.070-2.090)\rangle /\langle\mathcal{F}(2.235-2.255)\rangle}{\langle\mathcal{F}(2.235-2.255)\rangle /\langle\mathcal{F}(2.360-2.380)\rangle}.
\end{equation}
Here, $\langle \mathcal{F}(a-b)\rangle$ represents the median flux level in the region $[a, b]$, with both $a$ and $b$ in $\mu$m.
They also defined a relation between a star's metallicity, the H$_2$O--K2 index, and the equivalent width of the 2.21 $\mu$m sodium doublet and 2.26 $\mu$m calcium triplet. 
We calculated H$_2$O--K2 and the two equivalent widths, as well as their uncertainties, by creating a sequence of simulated spectra in which random noise is added to the observed flux consistent with the flux uncertainty at each wavelength.
We found the calculated H$_2$O--K2 values to be normally distributed such that H$_2$O--K2 $ = 0.919 \pm 0.002$. 
The equivalent width of the sodium doublet is $5.533 \pm 0.101$ \AA{} and the equivalent width of the calcium triplet is $3.863 \pm 0.089$ \AA.

\begin{figure}[htbp]
\centerline{\includegraphics[width=0.5\textwidth]{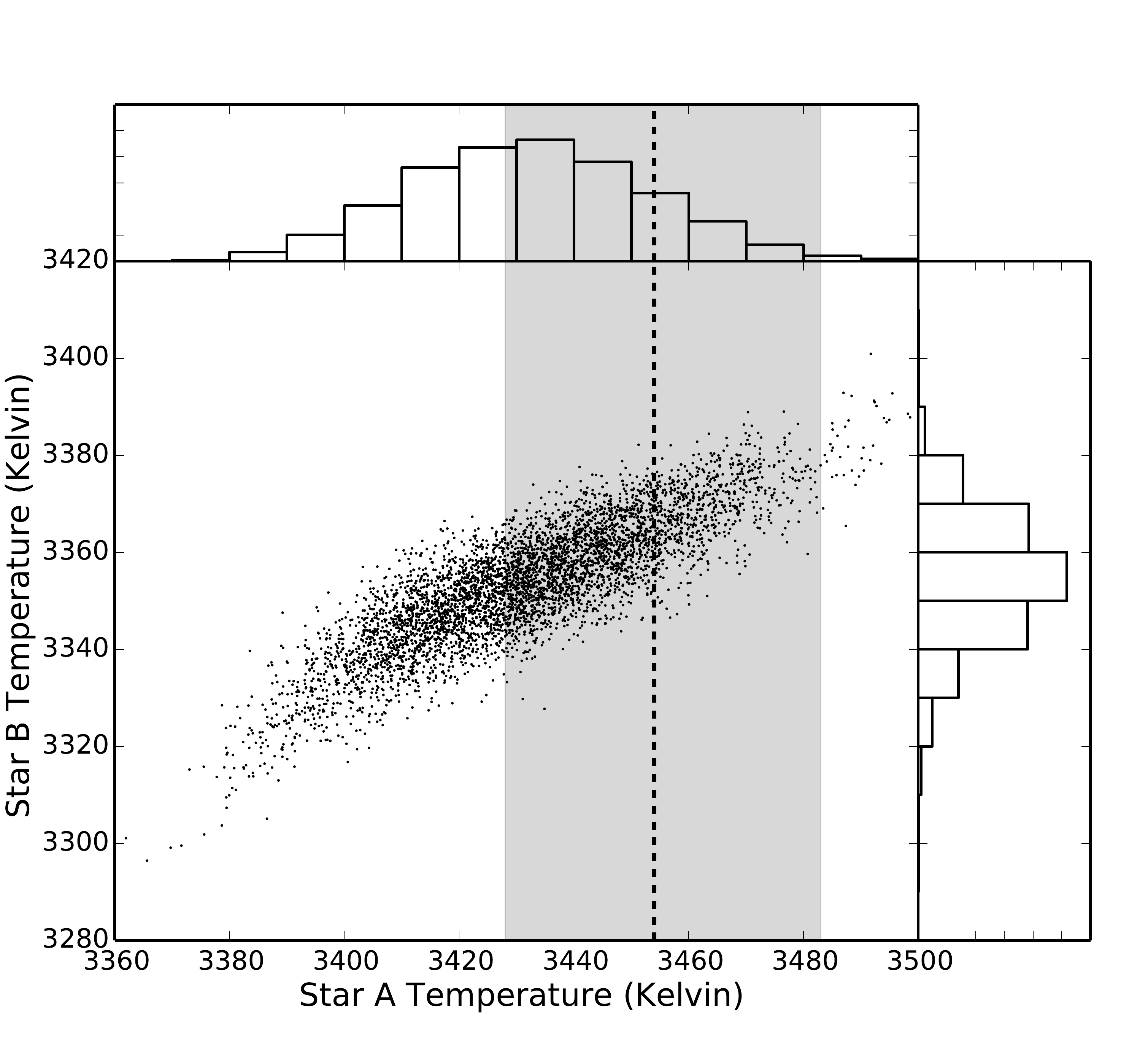}}
\caption{Joint posterior on the effective temperature of \LA{} and \LB. 
Marginalizing over the temperature of each star separately, we find the A component has a temperature of $3431 \pm 21$ K and the B component has a temperature of $3354 \pm 17$ K. The dashed line and shaded region correspond to the temperature of \LA{} expected based on our model-independent mass measurement from the combined transit and RV fit.
}
\label{TempPlot}
\end{figure}

If our spectrum consisted of the flux from only one star, we could convert our value directly into a stellar effective temperature and metallicity. 
In this case, each value is really the combination of two separate values, one for each M dwarf. 
However, if we assume the two stars have the same metallicity, useful information can still be extricated.
We first drew from the posterior of $\Delta K$ values from our PHARO near-infrared adaptive optics observations and our posteriors for H$_2$O--K2 and the equivalent widths. 
From these, we used the relations of \citet{RojasAyala12} to calculate the system metallicity. 
We then interpolated the table provided in that paper to find a relation between H$_2$O--K2 and effective temperature for a given metallicity. 
Using the Dartmouth stellar evolution models, we then determined which two modeled stars best fit both the observed flux ratio and combined H$_2$O--K2 index value. 
By repeating this process many times, continuously drawing from the posteriors for each measured value we determined a posterior on the temperature, and by extension the mass, of each star. 
The joint posterior on the temperature of the two stars is shown as Figure \ref{TempPlot}.

\subsection{Transit Parameters}

To measure the parameters of \LC, we forward modeled the \LA-C system over the timespan from the launch of \itk{} to the date of the final RV observation in 2013.
At each time corresponding to an RV observation, we calculated the expected radial velocity relative to a stationary \LB{} assuming a Keplerian orbit.
At each \itk{} timestamp during a transit or near the expected time of secondary eclipse, we calculated the expected relative flux assuming a \citet{Mandel02} light curve model. 
We fit four limb darkening parameters using the prescription of \citet{Claret11}, allowing the value for each limb darkening coefficient to float as a free parameter. 
In calculating the light curves, we used an adapted version of the PyAstronomy package\footnote{https://github.com/sczesla/PyAstronomy}, modified to allow eccentric orbits.

In all, we fit for 16 parameters: $\sqrt{e}\cos\omega$, $\sqrt{e}\sin\omega$, time of central transit, orbital period, brown dwarf mass, orbital inclination, \LA-C radius ratio, four limb darkening parameters, the third light from \LB, \logg{} of \LA, the secondary eclipse depth, the stellar mass, and the RV zeropoint (relative to \LB). We did not use an RV jitter term, as our RV uncertainties of $\sim 100$ m s$^{-1}$ are significantly larger than the jitter expected for a main-sequence M dwarf. We used {\tt emcee}, an affine-invariant ensemble sampler described by \citet{Goodman10} and implemented by \citet{Foreman-Mackey12}, to maximize the likelihood function
\begin{eqnarray}
\mathcal{L} &=& 0.5 \bigg[\sum_i \bigg(\frac{\textrm{RV$_{\textrm{model, $i$}}$} - \textrm{RV$_{\textrm{observed, $i$}}$}}{\sigma_{\textrm{RV}, i}}\bigg)^2 \nonumber \\
&+& \sum_i
 \bigg(\frac{f_{\textrm{model SC, $i$}} - f_{\textrm{observed SC, $i$}}}{\sigma_{f _\textrm{SC}, i}}\bigg)^2 \nonumber \\
 &+& \sum_i
 \bigg(\frac{f_{\textrm{model LC, $i$}} - f_{\textrm{observed LC, $i$}}}{\sigma_{f _\textrm{LC}, i}}\bigg)^2\bigg].
\end{eqnarray}
Here, $f_\textrm{LC}$ corresponds to the observed flux in the \itk{} long cadence data and $f_\textrm{SC}$ corresponds to the short cadence data.
The period we fit and report here is the period observed in the frame of an observer at the barycenter of the solar system, not in the frame of the \LHS{} system.
That is, we do not correct for relativistic effects induced by the star system's systemic velocity.

We imposed two different priors on the stellar mass, reflecting various levels of trust in theoretical stellar evolutionary models. 
First we apply the stellar empirical mass-radius relation of \citet{Boyajian12}, which encodes no direct model-dependent information, as a prior
We use their relation for ``single stars.''
While our star has a wide binary companion at tens of AU, the single collection is more representative of \LA{} than the short-period eclipsing binaries used to build the eclipsing binary main sequence of \citet{Boyajian12}.
Given a precise measurement of the stellar density \rhostar, semimajor axis $a/$\rstar, Doppler semiamplitude $K$, eccentricity, and inclination, the mass and radius of both the primary and secondary star can then be calculated.
We derive these relations in the appendix.

We next repeated this analysis, applying a prior on the stellar mass using the spectroscopic parameters from our TripleSpec analysis, as described in \textsection{\ref{TripleSpec}}.

In each of these cases, we can calculate the mass and radius of \LB{} through the Dartmouth models by comparing the relative brightness of \LA{} and \LB{} in conjunction with the (now known) mass of \LA. 
We can also measure a model-dependent distance to the system, which depends both on our measured mass and the mass-luminosity relation encoded in the stellar models.

\begin{figure}[htbp]
\centerline{\includegraphics[width=0.5\textwidth]{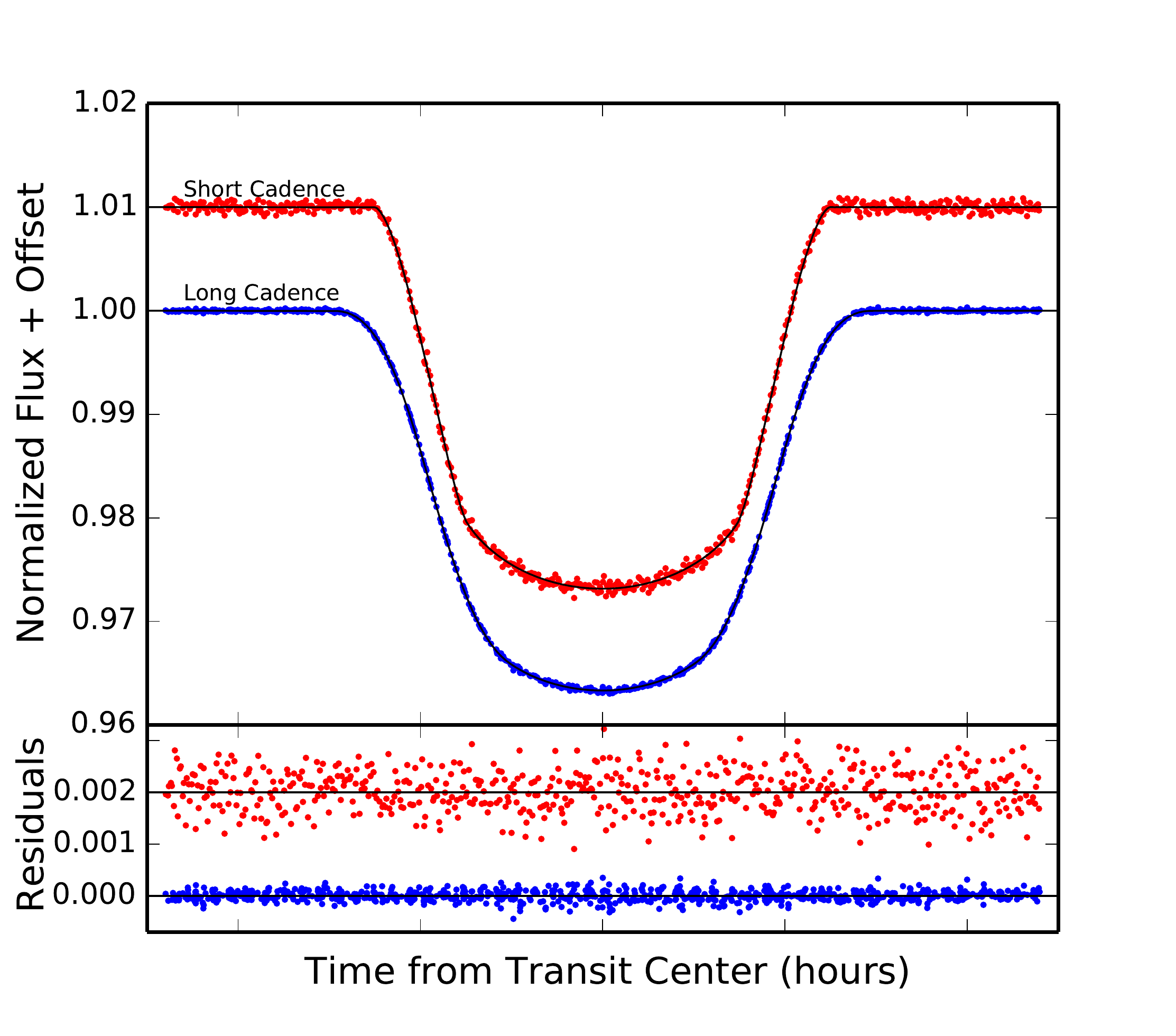}}
\caption{Phase-folded transit light curve, fit to the maximum likelihood model.
Blue points represent long cadence data and red points represent short cadence data.
The scale of the residuals is a factor of five larger than the scale of the light curve.
  }
\label{LC}
\end{figure}

\begin{figure}[htbp]
\centerline{\includegraphics[width=0.5\textwidth]{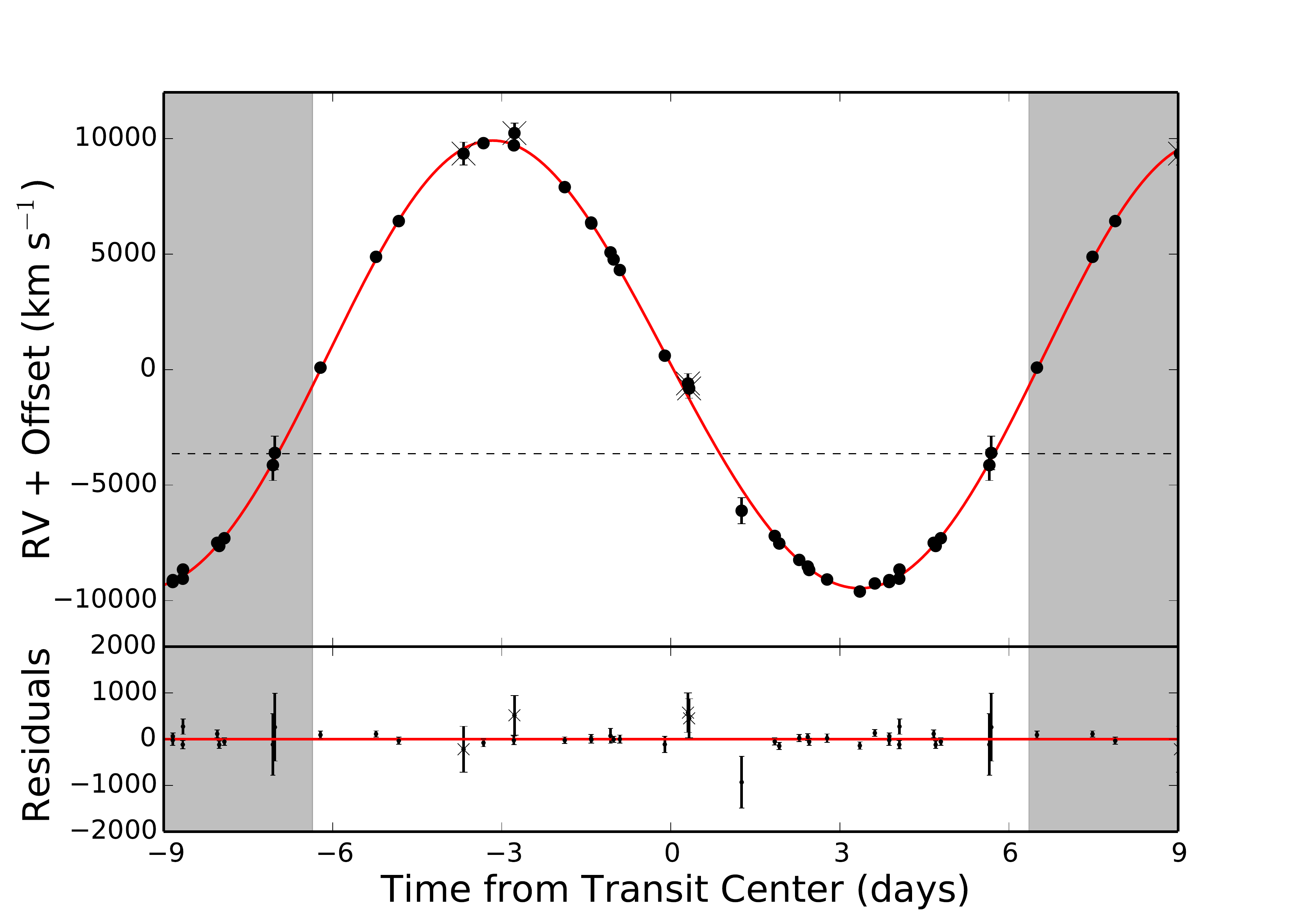}}
\caption{Phase-folded RV data curve, fit to the maximum likelihood model.
For the majority of observations, the data points are larger than the size of the error bars.
The gray shaded regions represent an extension of the RV data beyond one phase to provide clarity for the reader. 
Observations marked with an cross represent data collected while using the iodine cell.
The dashed line represents the RV of \LB{}, which does not change at the level of our precision over the 3-year RV baseline.
  }
\label{RVs}
\end{figure}

The best fit model to the light curve data and RVs are plotted in Figures \ref{LC} and \ref{RVs}, respectively.

\section{Results}

The orbital parameters for \LC{} are listed in Table \ref{OrbitTable}.
The physical properties of the \LHS{} system are listed in Table \ref{BigTable}. 
In the latter table, we include two columns of values. 
The first set of values represents the values we find using our data-driven model, using only the empirical mass-radius relation of \citet{Boyajian12} without any direct use of stellar models. 
The second set of values corresponds to the inclusion of a model-dependent prior on the stellar mass.
In this case, we impose as a prior our mass derived from the near-IR spectroscopy, found in \textsection{\ref{TripleSpec}}.

We find that we are able to measure the observed transit depth, uncorrected for the third light contributions of \LB, to a precision of 0.5\%.
We are additionally able to measure the Doppler semiamplitude $K$ to 0.3\%. 
Therefore, our uncertainties in the brown dwarf's physical parameters are dominated by the uncertainties on the absolute physical parameters of the two M dwarfs in the system.

We can measure the stellar mass directly from the light curve and RV observations without any direct reliance on theoretical stellar models, as shown in the Appendix.
In this case, we measure a mass for \LA{} of $0.381 \pm 0.019$ \msun{} and a radius of $0.380 \pm 0.007$ \rsun.  
We then find a mass and radius of \LC{} of $64.6 \pm 2.1$ \mjup{} and $0.798 \pm 0.014$ \rjup, respectively.
Thus, in this case we can measure the mass of the brown dwarf to a precision of $3.2\%$ and the radius to $1.8\%$. 

From our near-IR spectroscopic analysis of the system, we measure a temperature for \LA{} of $3431 \pm 21$ K, which gives us a mass of $0.339 \pm 0.016$ \msun. 
We repeat our analysis, using this value as a prior on our stellar mass.
In this case, we find a value for the stellar mass between our empirical value and that imposed by our model prior: $0.358 \pm 0.011$ \msun.
We then find a mass for the brown dwarf of $62.1 \pm 1.2$ \mjup{} and a radius of $0.782 \pm 0.013$ \rjup. 
This is a model-dependent mass measured to a precision of $1.9\%$ and a model-dependent radius to $1.4\%$. 

Our brown dwarf mass is consistent with that found by J11, while our radius is smaller at the $1.4\sigma$ level. 
Part of this discrepancy may be due to the choice of models used: these authors used the Padova model grids of \citet{Girardi02}. 
These models predict a larger mass than both the Dartmouth models we use and the BT-Settl models \citep{Allard10}.
Using the Padova models, the authors of the discovery paper adopted a slightly smaller \logg, which for a given mass implies a larger star, and therefore a larger planet.
The discrepancy may also be affected by our choices of limb darkening models: the authors of the discovery paper use a quadratic limb darkening model.
With only five transits observed, this is a reasonable choice. 
Given the signal to noise obtained from fitting four years of \itk{} data simultaneously, we require a four-parameter limb darkening solution to develop an appropriate model fit.

Our mean density for \LC{} is 40\% larger than that reported in the discovery paper.
This appears to be because the authors of that paper misreported their density, as it is inconsistent with their reported mass and radius. 
These authors may have reported the density relative to Jupiter, not in units of g cc$^{-1}$ as listed in their Table 5.
Even with this correction, the density we report is larger than the density of J11 due to the difference in the radius of the brown dwarf described in the previous paragraph.

We measure a period of $12.7137941 \pm 0.0000002$ days in the frame of the solar system. 
The uncertainty in the period is 17 milliseconds, and the period is measured to a precision of 15 parts per billion.

We measure the total mass in the \LA C system to a precision of 4.8 percent. 
Neglecting our uncertainty in the measured period, from differentiating Kepler's Third Law we expect our measurement of the semimajor axis to be three times more precise than that of the total mass.
In fact, we measure a semimajor axis of $0.0812 \pm 0.0013$ AU, a precision of 1.6 percent.

\subsection{Secondary Eclipse Observation}

J11 do not detect a secondary eclipse and can only place an upper limit of 65 parts per million on the potential eclipse depth.
With a full four years of \itk{} data, we are considerably more sensitive to eclipses.
From the RVs and shape of the primary eclipse alone, we know the A-C system has a nonzero eccentricity: we find $e \cos \omega = 0.024 \pm 0.003$.
As a result, we expect the secondary eclipse to occur approximately 4.5 hours after the midpoint between consecutive primary transits. 

\begin{figure}[htbp]
\centerline{\includegraphics[width=0.5\textwidth]{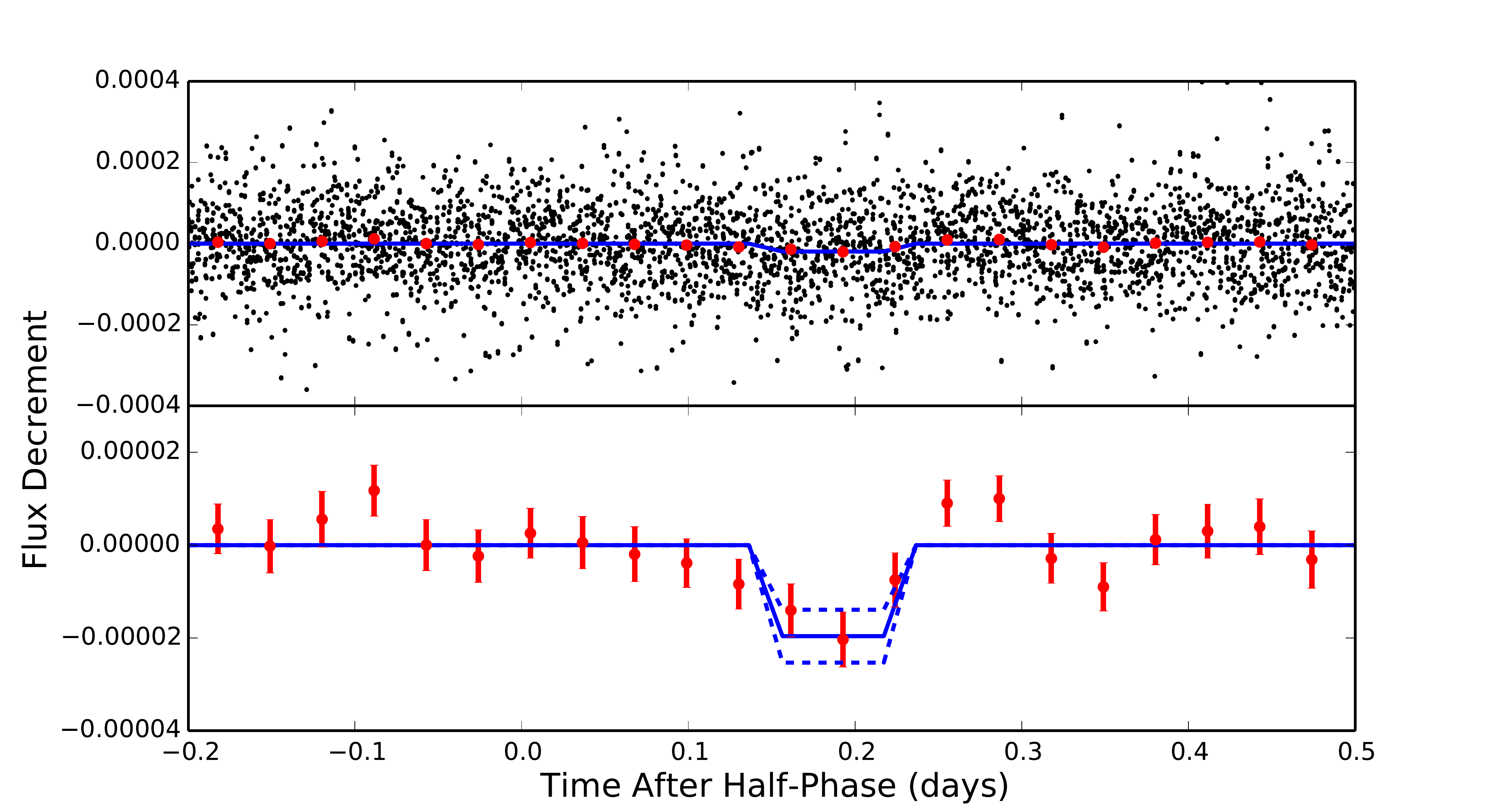}}
\caption{Secondary eclipse of \LC{} as observed by \itk. 
(top) In black, the \itk{} data are phase-folded and plotted; we bin every 0.03 days of observations together to reduce the apparent scatter, as shown in red. 
As the noise is nearly completely white, this is justified for plotting purposes.
In blue is our best-fitting secondary eclipse model.
We treat the brown dwarf as a uniform sphere in our modeling efforts.
(bottom) Same as the above, excluding the raw data. 
We detect an eclipse depth of $25 \pm 7$ ppm after accounting for the correction for the third light contribution from \LB.
The dashed blue lines represent the $1\sigma$ deviation in eclipse depth from the best-fitting model.
}
\label{SecondaryPlot}
\end{figure}

When we include a secondary eclipse in our system model, we detect a signal at $3.5\sigma$, as shown in Figure \ref{SecondaryPlot}.
This eclipse has a depth of $25 \pm 7$ parts per million and occurs $4.44 \pm 0.16$ hours after the midpoint between primary transits.
From these data, we measure $e \cos \omega = 0.0228 \pm 0.0008$.

\subsection{Distance to the \LHS{} System}

There is, at present, no measured parallax to the \LC{} system. 
We must therefore rely on stellar models to convert the measured apparent magnitudes to distance estimates.
J11, using the Padova model atmospheres, announced a distance to the system of $36.6 \pm 1.1$ pc. 
The Dartmouth models predict a lower mass, and therefore a lower luminosity for \LA{}, so to maintain the observed brightness of the system from $g$ to $K_s$-band, these models require a smaller distance modulus.
We find a model-dependent distance to the system of $32.7 \pm 1.3$ pc.
A measured parallax to this system, either from the ground or from Gaia, will be useful for resolving the $2\sigma$ discrepancy between these distances, informing the upcoming next generation of stellar evolution models.

\begin{deluxetable*}{lccc}
\tablecaption{Orbital Parameters for the LHS\,6343\,AC System}
\footnotesize
\tablewidth{0pt}
\tablehead{
  \colhead{Parameter} & 
  \colhead{Value}     &
  \colhead{} &
  \colhead{$1\sigma$ Confidence}  \\
  \colhead{} & 
  \colhead{} &
  \colhead{} &
  \colhead{Interval}      
}
\startdata
Orbital Period, $P$~[days] & 12.7137941 & $\pm$& 0.0000002 \\
Transit Center (TDB $- 2440000$) & 15008.07259 & $\pm$ &  0.00001 \\
Radius Ratio, $(R_P/R_\star)$ &      0.216 & $\pm$ &       0.004 \\
Observed Transit Depth (percent) &  3.198 & $\pm$ &       0.015 \\
Scaled Semimajor axis, $a/R_\star$  &  46.0 & $\pm$ &       0.4 \\
Orbital Inclination, $i$~[deg] &  90.45 & $\pm$ &     0.03  \\
Transit Impact Parameter, $b$ &  0.36 & $\pm$ &     0.02 \\
Argument of Periastron $\omega$~[degrees] & -40 & $\pm$ &      4 \\
Eccentricity  &  0.030 & $\pm$ &      0.002  \\
Secondary Phase ($e \cos \omega$) & 0.0228 & $\pm$ &      0.0008 \\
Secondary Depth (ppm) & 25 & $\pm$ & 7 \\
Velocity semiamplitude $K_A$~[km s$^{-1}$] &  9.69 & $\pm$ &       0.02 \\
Star A-B RV Offset [km s$^{-1}$] & 3.64 & $\pm$ &       0.02  
\enddata
\tablecomments{All parameters calculated by simultaneously fitting to the RV data and \itk{} data near the times of transit and secondary eclipse.}
\label{OrbitTable}
\end{deluxetable*}

\begin{deluxetable*}{lccccccc}

\tablecaption{Physical Parameters for LHS\,6343\,ABC}
\footnotesize
\tablewidth{0pt}
\tablehead{
  \colhead{Parameter} & 
  \colhead{Value}     &
  \colhead{} &
  \colhead{$1\sigma$ Confidence}     &
  \colhead{Value}     &
  \colhead{} &
  \colhead{$1\sigma$ Confidence}     &
  \colhead{Comment}   \\
  \colhead{} & 
  \colhead{(Empirical Prior)}     &
  \colhead{} &
  \colhead{Interval}     & 
  \colhead{(Model Prior)} & 
  \colhead{} & 
  \colhead{Interval}     &
  \colhead{}  
}
\startdata
\emph{Stellar Parameters} & & & \\
$M_A$~[\msun] &  0.381 & $\pm$ &       0.019 &      0.358 & $\pm$ &      0.011 & A \\
$M_B$~[$M_\odot$] & & & & 0.292 & $\pm$ & 0.013 & A \\
$R_A$~[$R_\odot$] &   0.380 & $\pm$ &       0.007 &      0.373 & $\pm$ &      0.005 & A \\
$R_B$~[$R_\odot$] & & & & 0.394 & $\pm$ & 0.012 & A \\
$\rho_A$~[$\rho_\odot$] & 6.96 & $\pm$ &       0.19 &      6.93 & $\pm$ &      0.19 & A \\
$\log g_A$~[cgs] &  4.86 & $\pm$ &        0.01 &       4.85 & $\pm$ &       0.01 & A \\
Metallicity [Fe/H] & & & & 0.03 & $\pm$ & 0.26 & B \\
Metal Content [a/H]  & & & & 0.02 & $\pm$ & 0.19 & B \\
Distance~[pc] & & & & 32.7 & $\pm$ & 1.3 & C \\
Flux Ratio $F_B/F_A, K_p$ &   0.461 & $\pm$ &       0.055 &      0.518 & $\pm$ &    0.032 & A \\
$\Delta K_p$ [magnitudes] & 0.84 & $\pm$ & 0.12 & 0.71 & $\pm$ & 0.07 & A \\
$T_{{\rm eff}, A}$ [K] & & & & 3431 & $\pm$ & 21  & B\\
$T_{{\rm eff}, B}$ [K] & & & & 3354 & $\pm$ & 17  & B\\
 & & \\
\emph{Brown Dwarf Parameters} & & & \\
$M_C$~[\mjup] & 64.6 & $\pm$ &        2.1 &      62.1 & $\pm$ &       1.2 & A \\
$R_C$~[\rjup] & 0.798 & $\pm$ &       0.014 &      0.783 & $\pm$ &      0.011  & A\\
Semimajor Axis, A-C System (AU) &  0.0812 & $\pm$ &      0.0013 &     0.0797 & $\pm$ &     0.0008 & A \\
Mean Planet Density, $\rho_C$~[g cm$^{-3}$] & 170 & $\pm$ &       5. &    173 & $\pm$ &      5 & A \\
$\log g_C$~[cgs] &  5.419 & $\pm$ &       0.008 &      5.420 & $\pm$ &      0.008 & A \\
$T_{\rm eq}$ ($T_{\rm eff}(\frac{R_\star}{2a})^{1/2}$)~[K] & & & & 358 & $\pm$ & 3 & A,B
\enddata

\tablecomments{(A) Calculated by simultaneously fitting to the RV data and \itk{} data near the times of transit and secondary eclipse. \\
(B) Measured from near-IR spectroscopy following the method of \citet{RojasAyala12}. \\
(C) Calculated by fitting the observed apparent magnitudes to model-predicted absolute magnitudes.
}
\label{BigTable}
\end{deluxetable*}

\section{Discussion}

\begin{figure}[htbp]
\centerline{\includegraphics[width=0.5\textwidth]{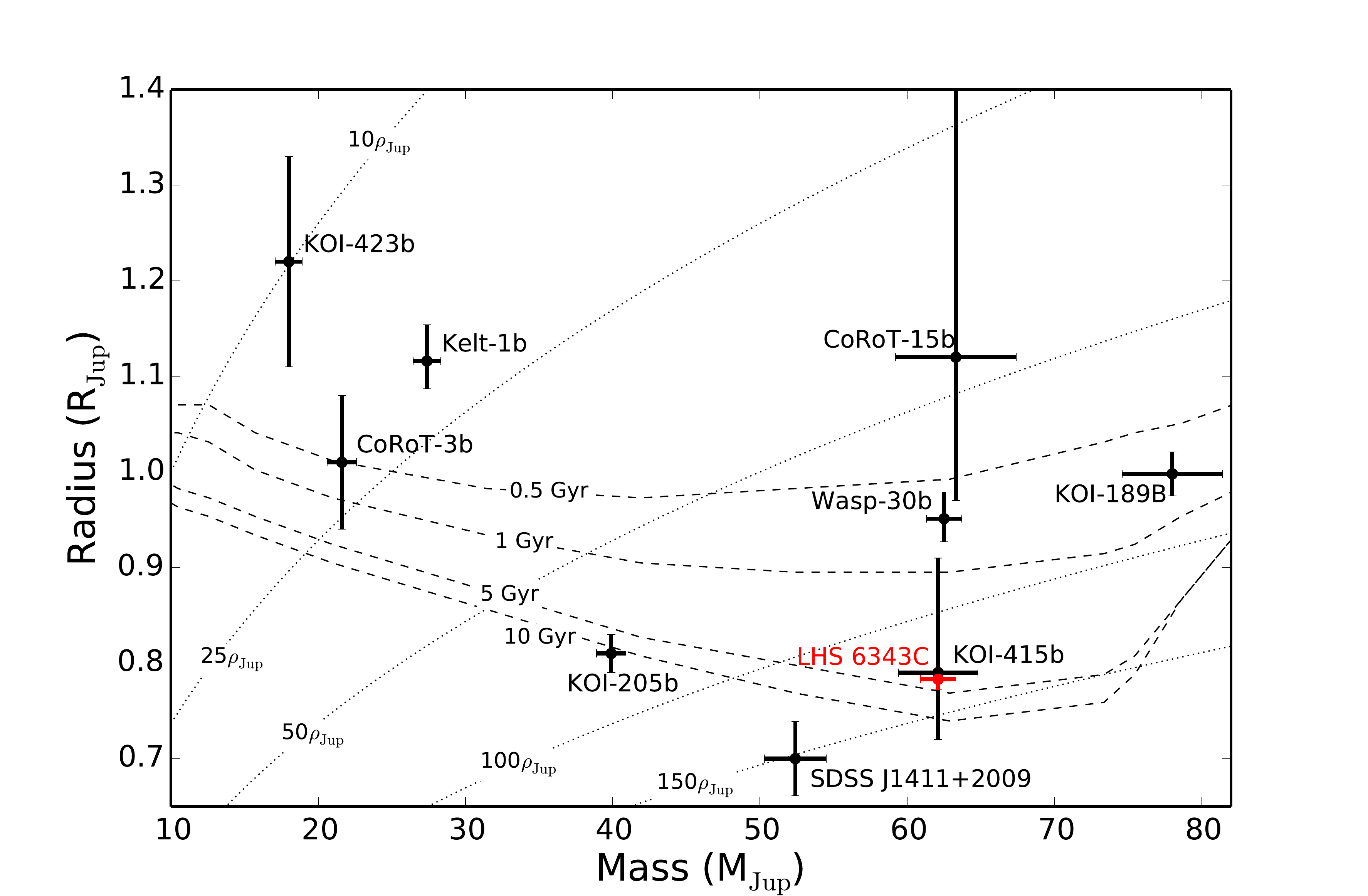}}
\caption{Mass-radius diagram for known transiting brown dwarfs.
The dashed lines represent the \citet{Baraffe03} isochrones for (top to bottom) ages of 0.5, 1.0, 5.0, and 1.0 billion years.
The dotted lines are isodensity contours for (top left to bottom right) densities corresponding to 10, 25, 50, 100, and 150 times the density of Jupiter. 
\LC{} has a density of $130 \pm 4 \rho_\textrm{Jup}$ and appears to have an age of 3-5 Gyr.
Data taken from \citet{Deleuil08, Bouchy11a, Bouchy11b, Siverd12, Diaz13, Moutou13, Triaud13, Diaz14, Littlefair14}.
Not shown are the components of the young binary brown dwarf system
2MASS 2053-05 \citep{Stassun06}, which have
radii well above the plot range.
  }
\label{MassRadiusPlot}
\end{figure}

There are now nine brown dwarfs with measured masses and radii \citep{Moutou13}. 
Of this sample, there are only four that are not inflated due to youth or irradiation.
\LC{} is effectively a field brown dwarf: the equilibrium temperature for an object at its orbital separation is 360 K while a 65 \mjup{} brown dwarf is expected to cool to only 700 K over a Hubble time \citep{Burrows01}.
Thus, the irradiation from the primary star on the brown dwarf is negligible.
Additionally, since the system has a nonzero eccentricity, the system is not tidally locked, minimizing any effects the primary star may have on any one point on the brown dwarf's surface.
\LC{} can be used as a laboratory to study the physics of solitary brown dwarfs, as it is effectively a field brown dwarf with a known mass, radius, and metallicity.
The sample of transiting brown dwarfs that can be used to probe the physics of field brown dwarfs is highly limited, making each individual system extremely valuable.

There is some evidence that our current best understanding of the physics of brown dwarfs is incomplete.
\citet{Dupuy09} find evidence for a ``substellar luminosity problem,'' in which the brown dwarf binary HD\,130948\,BC is twice as luminous as predicted by evolutionary models.
A similar result is found in the Gl 417 BC system \citep{Dupuy14}.
As these are the only two brown dwarf systems with reliable measurements of both mass and age, this result is suggestive of a fundamental issue with substellar models.

We have only a lower limit on the age of the system: J11 find no youth indicators present in the \LHS{} system so it is likely not less than 1-2 Gyr old.
Therefore, a measured luminosity would be most useful as a probe of this specific plane if the luminosity were consistent with extreme youth ($< 1$ Gyr) or extreme age ($>14$ Gyr).
A measured luminosity is still useful, as it allows us to locate the brown dwarf's position in the mass-radius-luminosity plane. 
While there is a collection of non-inflated brown dwarfs with masses and luminosities measured, there are only three with mass and radius and none with both radius and luminosity.
Moreover, we also know the metallicity of the brown dwarf, assuming it has the same composition as \LA B. 

There is a degeneracy between the inferred age of the system and the atmosphere of the brown dwarf.
Specifically, a brown dwarf with the mass and radius of \LC{} would be expected to be significantly older if it were covered with optically-thick clouds, as the clouds would keep the brown dwarf at a hotter internal adiabat. 
The models of \citet{Baraffe03}, which do not include clouds, suggest an age of approximately 5 Gyr, consistent with the cloudless models of \citet{Saumon08}.
However, \citet{Saumon08} predict a cloudy brown dwarf with a mass of \LC{} and an age equal to the age of the universe would have a radius $2\sigma$ larger than that observed for this object. 
This is consistent with the models of \citet{Burrows11}, who find the system must be very old if \LC{} has a thick layer of clouds.
These authors claim thinner clouds or no clouds may be preferred by the data.
Therefore, any additional observations which suggest the presence of clouds on \LC{} would be at odds with the predictions from theoretical brown dwarf model atmospheres.

The luminosity of \LC{} can be measured by observing its secondary eclipses as it passes behind \LA.
In the \itk{} bandpass, we find the eclipse depth is $25 \pm 7$ parts per million. 
Between 1 and 3 microns, the depth is expected to be 0.1\%, observable with ground-based telescopes.
In the 4.6 $\mu$m Spitzer bandpass, the eclipse depth may be as large as 0.5$\%$ if the brown dwarf's atmosphere is cloud-free. 
We will observe this system during four secondary eclipse events in Spitzer Cycle 10, observing two eclipses in each available IRAC bandpass.
In addition to probing for extreme variability caused by patchy clouds in the atmosphere of \LC, combining these observations with the \itk{} secondary and ground-based JHK photometry will enable us to measure a luminosity of this brown dwarf from the visible to the mid-infrared. 
These observations will allow us to place the first data point on the brown dwarf mass-radius-metallicity-luminosity plane, testing the underconstrained brown dwarf atmospheric models in this parameter space for the first time.

\acknowledgements
We thank Luan Ghezzi and Jennifer Yee for helpful discussions which improved the quality of this manuscript.

The RV data presented herein were obtained at the W.M. Keck Observatory, which is operated as a scientific partnership among the California Institute of Technology, the University of California and the National Aeronautics and Space Administration.
The Observatory was made possible by the generous financial support of the W.M. Keck Foundation.
We are grateful to the entire Kepler team, past and present. 
Their tireless efforts were all essential to the tremendous success of the mission and the future successes of K2.

Some of the data presented in this paper were obtained from the Mikulski Archive for Space Telescopes (MAST). STScI is operated by the Association of Universities for Research in Astronomy, Inc., under NASA contract NAS5--26555. Support for MAST for non--HST data is provided by the NASA Office of Space Science via grant NNX13AC07G and by other grants and contracts. This paper includes data collected by the \itk{} mission. Funding for the \itk{} mission is provided by the NASA Science Mission directorate.

B.T.M. is supported by the National Science Foundation Graduate Research Fellowship under Grant No. DGE‐1144469. 
J.A.J. is supported by generous grants from the David and Lucile Packard Foundation and the Alfred P. Sloan Foundation.
C.B. acknowledges support from the Alfred P. Sloan Foundation.

The Robo-AO system is supported by collaborating partner institutions, the California Institute of Technology and the Inter-University Centre for Astronomy and Astrophysics, and by the National Science Foundation under Grant Nos. AST-0906060, AST-0960343, and AST-1207891, by the Mount Cuba Astronomical Foundation, by a gift from Samuel Oschin.

We would like to thank the staff of both Palomar Observatory and the W.M. Keck Observatory for their support during our observing runs. 
Finally, we wish to acknowledge and recognize the very significant cultural role and reverence that the summit of Maunakea has always had within the indigenous Hawaiian community. 
We are most fortunate to have the ability to conduct observations from this mountain.

{\it Facilities:} \facility{Keck:I (HIRES)}, \facility{Kepler}, \facility{PO:Hale (TripleSpec)}, \facility{PO:1.5m (Robo-AO)}

\appendix
\section{Derivation of Direct Mass and Radius Measurement}
\label{Derivation}

\citet{Seager03} derive four directly observable parameters in an exoplanet light curve under a specific set of assumptions. 
Namely, they assume circular orbits, $M_2 \ll M_1$, and that the third light contribution from a blended star is zero.
None of these are true for the \LHS{} system. 
As a result, the derivation which follows provides an analytic result which is exactly true when written in terms of physical parameters, but when common approximations for these parameters in terms of observables such as the transit duration, impact parameter, and relative flux decrement during transit are substituted for these parameters, the results below only approximate the truth. 
When calculating physical parameters using this method, care should be taken to avoid using these oversimplified expressions.

Following \citet{Seager03}, the transit light curve enables a direct measurement of the stellar density \rhostar{} and the reduced semimajor axis and the stellar radius, $a/$\rstar. 
From these, the authors claim if the stellar mass-radius relation is known, then the stellar mass can be measured directly from the light curve. 
We show if the Doppler semiamplitude $K$ is known, the stellar mass can be measured exactly.

We know from Kepler's Third Law that, for two orbiting bodies with masses \mstar{} and $m_p$ (by convention, \mstar$ > m_p$) and orbital period $P$, that
\begin{equation}
a = \bigg(\frac{GP^2(M_\star+m_p)}{4\pi^2}\bigg)^{1/3},
\end{equation}
where $G$ is Newton's constant.
The mean stellar density is defined for a star of mass \mstar{} and radius \rstar{} to be
\begin{equation}
\rho_\star = \frac{3M_\star}{4 \pi R_\star^3}.
\label{Eq:Density}
\end{equation}

We can combine these two in such a way that we recover an expression for the mass ratio that depends only on observable parameters. We find
\begin{equation}
1 + \frac{m_p}{M_\star} = \bigg(\frac{3\pi}{GP^2}\bigg) \bigg(\frac{1}{\rho_\star}\bigg) \bigg(\frac{a}{R_\star}\bigg)^3 \equiv c_1.
\end{equation}

Famously, the Doppler semiamplitude $K$ observed in a radial velocity survey is 
\begin{equation}
K = \bigg(\frac{2\pi G}{P}\bigg)^{1/3} \frac{m_p \sin{i}}{(M_\star + m_p)^{2/3}}\frac{1}{\sqrt{1-e^2}}.
\end{equation}
Here, $i$ is the orbital inclination and $e$ the eccentricity, while all other variables retain their previous meaning.
Rearranging this equation, we can once again write the mass ratio in terms of observable parameters only. In this case,
\begin{equation}
\frac{m_p^3}{(M_\star + m_p)^2} = \frac{K^3 P}{2\pi G} \bm{\bigg(}\frac{\sqrt{1-e^2}}{\sin{i}}\bm{\bigg)^3} \equiv c_2
\end{equation}

With two equations and two unknown masses, we can solve for the primary and secondary mass individually. We find
\begin{align}
M_\star &= \frac{c_1^2 c_2}{(c_1-1)^3}  \nonumber \\
        &= \frac{\bigg(\frac{9\pi}{2}\bigg)\bigg(\frac{1}{\rho_\star}\bigg)^2\bigg(\frac{a}{R_\star}\bigg)^6\bigg(\frac{K}{GP}\bigg)^3\bigg(\frac{\sqrt{1-e^2}}{\sin{i}}\bigg)^3}{\bigg[\bigg(\frac{3\pi}{GP^2}\bigg) \bigg(\frac{1}{\rho_\star}\bigg) \bigg(\frac{a}{R_\star}\bigg)^3-1\bigg]^3} \\
\end{align}
and
\begin{align}
m_p     &= \frac{c_1^2 c_2}{(c_1-1)^2}  \nonumber \\
        &= \frac{\bigg(\frac{9\pi}{2}\bigg)\bigg(\frac{1}{\rho_\star}\bigg)^2\bigg(\frac{a}{R_\star}\bigg)^6\bigg(\frac{K}{GP}\bigg)^3\bigg(\frac{\sqrt{1-e^2}}{\sin{i}}\bigg)^3}{\bigg[\bigg(\frac{3\pi}{GP^2}\bigg) \bigg(\frac{1}{\rho_\star}\bigg) \bigg(\frac{a}{R_\star}\bigg)^3-1\bigg]^2} \\
\end{align}

From the stellar density, the calculated mass can be used to measure the stellar radius. 
Plugging this equality in to Equation \ref{Eq:Density} above, we find that
\begin{equation}
R_\star = \frac{\bigg(\frac{3}{2}\bigg)\bigg(\frac{1}{\rho_\star}\bigg)\bigg(\frac{a}{R_\star}\bigg)^2\bigg(\frac{K}{GP}\bigg)\bigg(\frac{\sqrt{1-e^2}}{\sin{i}}\bigg)}{\bigg[\bigg(\frac{3\pi}{GP^2}\bigg) \bigg(\frac{1}{\rho_\star}\bigg) \bigg(\frac{a}{R_\star}\bigg)^3-1\bigg]}.
\end{equation}

From a known stellar radius, the transit depth can be used to measure the planet radius directly. For a flux decrement $\Delta F$, 

\begin{equation}
R_p = R_\star\sqrt{\Delta F}.
\end{equation}

Therefore, by measuring the stellar density, reduced semimajor axis, orbital period, transit depth, inclination, eccentricity, and Doppler semiamplitude, we can measure the stellar and planetary mass and radius. 
Moreover, since the companion is transiting, we know $\sin i \approx 1$. 

\citet{Dawson12a} present equations for the physical parameters above in terms of parameters directly observable from the light curve. 
Specifically, they find, in the limit of $m_p << M_\star$,
\begin{equation}
\frac{a}{R_\star} = \frac{2 \delta^{1/4} P}{\pi \sqrt{T^2_{14} - T^2_{23}}} \frac{\sqrt{1-e^2}}{1+e \sin w} 
\end{equation}
and
\begin{equation}
\rho_\star = \bigg[\frac{2 \delta^{1/4}}{ \sqrt{T^2_{14} - T^2_{23}}}\bigg]^3 \bigg(\frac{3P}{G\pi^2}\bigg)\bigg(\frac{\sqrt{1-e^2}}{(1+e \sin w)}\bigg)^3.
\end{equation}
Here, $\delta = (R_p / R_\star)^2$ is the fractional transit depth, or the relative areas of the transiting companion and the host star. 
$T_{14}$ is the transit duration from first to fourth contact (including ingress and egress), and $T_{23}$ is the transit duration from second to third contact (excluding ingress and egress).
 
If we substitute these into our above equations for the stellar mass and radius, we find our expressions for the mass and radius are undefined. 
Specifically, our denominator, $c_1 - 1$ is undefined at $m = 0$. 
Our equations above work specifically in the case where the mass of the companion is not negligible. 
This is because the stellar density cannot be measured exactly from the light curve alone.
While often neglected in exoplanet studies, the true observable is $(M_\star + m_p)/R_\star^3$. 
In cases where the mass ratio is large, this value approaches $M_\star / R_\star^3$, enabling the stellar density to be approximated well.
For the case of a Jupiter-sized planet transiting a sun-like star, such an approximation is reasonable.
However, this approximation breaks down for small mass ratios.
In this case, an additional constraint is required.

An additional constraint can be provided by using the mass ratio, which can be measured by observing ellipsoidal variations in the full phase curve \citep{Loeb03}. 
Ellipsoidal variations have been used both to confirm transiting planets \citep[e.g.][]{Mislis12} and to measure the mass ratios of already-confirmed planets \citep[e.g.][]{Welsh10, Jackson12}. 
By including such an observation, the degeneracy between the stellar density and mass ratio can be broken and the stellar mass measured directly.

\begin{figure}[htbp]
\centerline{\includegraphics[width=0.5\textwidth]{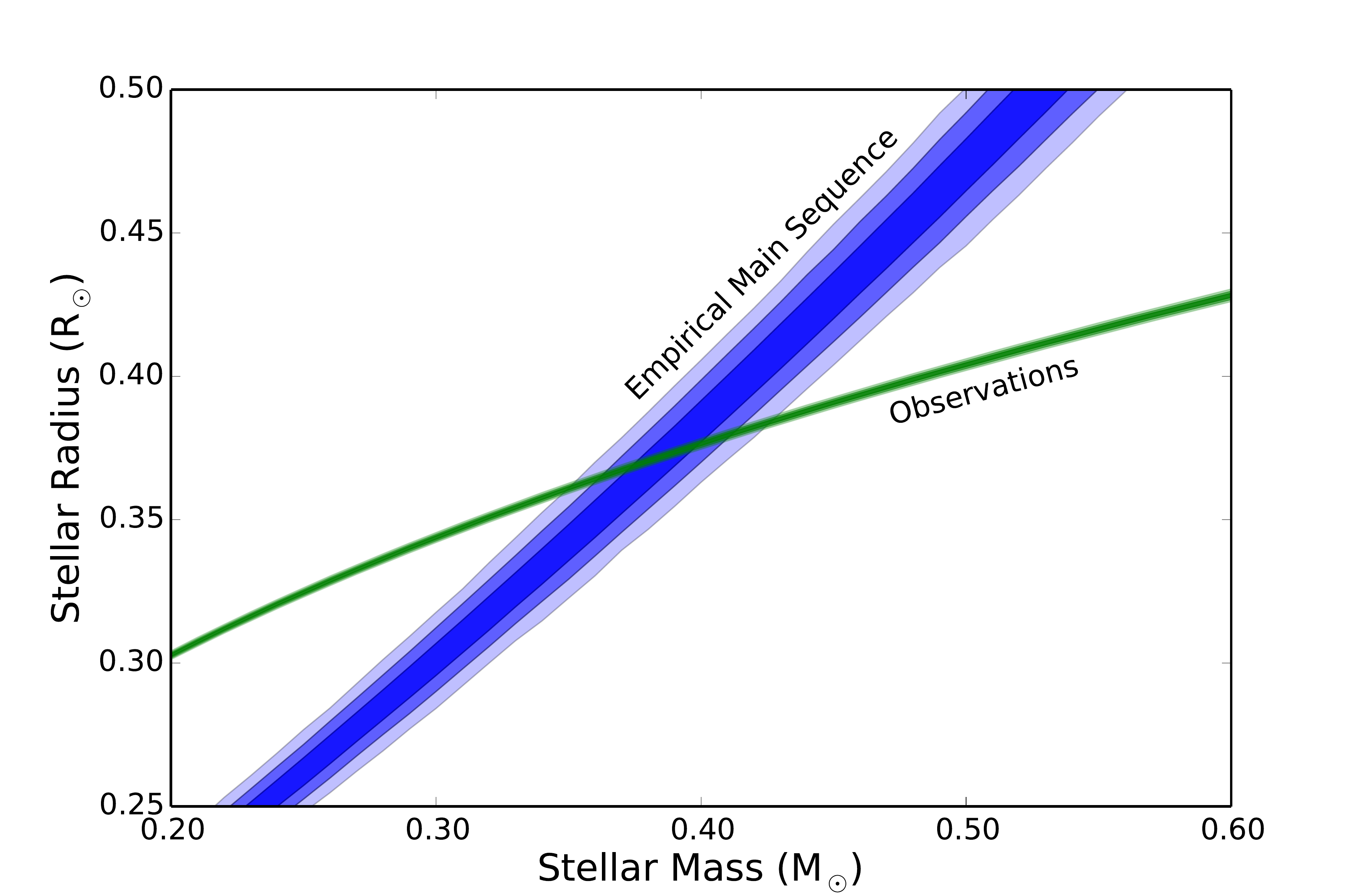}}
\caption{(green) Mass-radius relation for \LA{} from the observed transit light curve and RV observations, plotted with (blue) the mass-radius relation for K and M dwarfs of \citet{Boyajian12}. There are many possible stellar masses and radii which are formally allowed, but are unphysical. By combining weak constraints from empirical observations of the main sequence, a robust direct measurement on the mass and radius of both \LA{} and \LC{} can be made.
}
\label{M-R}
\end{figure}

When both the mass ratio is small and ellipsoidal variations cannot be observed from the light curve, the masses can still be measured directly if the star can be assumed to fall on the main sequence, as outlined by \citet{Seager03}. 
For a fixed transit depth, reduced semimajor axis, and Doppler semiamplitude, a star's inferred mass is related to the star's predicted radius such that $M \propto R^{>3}$, with the exact coefficient depending on the host-companion mass ratio (and approaching 3 as the mass ratio becomes infinite). Since the stellar main-sequence has a significantly different mass-radius relation, this information can be used to rule out many unphysical transit models. An example of this is shown as Fig. \ref{M-R}.

Because a nonzero mass ratio is required, this method is likely only applicable when the companion is a hot Jupiter, transiting brown dwarf, or low-mass stellar companion.
Moreover, it requires precise knowledge of both the Doppler semiamplitude and transit parameters. 
Therefore, the potential of this method is likely limited at present to hot transiting companions orbiting bright host stars.
Yet for these cases this technique may be very useful, especially when stellar evolutionary models may have systematic errors, such as when the host is an M dwarf or subgiant star.

\end{document}